\documentclass{PoS}
\usepackage{mathtools}
\usepackage{cancel}
\usepackage{comment}
\usepackage{capt-of}
\title{Energy loss of $B$ and $D$ mesons in PbPb collisions at $\sqrt{s_{NN}}$ = 2.76 TeV}

\ShortTitle{Heavy Quark Energy Loss}

\author{\speaker{Kapil Saraswat}\\
        Department of Physics \\
        Banaras Hindu University \\
        Varanasi, India \\
        E-mail: \email{kapilsaraswat76@gmail.com}}

\author{Prashant Shukla\\
        Nuclear Physics Division\\
        Bhabha Atomic Research Center  \\
        Mumbai, India \\
        E-mail: \email{pshuklabarc@gmail.com}}

\author{Venktesh Singh\\
        Department of Physics \\
        Banaras Hindu University \\
        Varanasi, India \\
        E-mail: \email{venkaz@yahoo.com}}

\abstract{
  We present the calculations of collisional and radiative energy loss of 
$B$ and $D$ mesons in the medium produced in PbPb collisions at $\sqrt{s_{NN}}$ 
= 2.76 TeV. The nuclear modification factor $R_{AA}$ of $B$ and $D$ mesons 
including shadowing and energy loss are calculated and compared with the 
measured data. While the $D$ meson $R_{AA}$ can be described in terms of 
the radiative energy loss alone, both the collisional as well as radiative 
energy loss are required to explain the $B$ meson measurements.

}

\FullConference{7th International Conference on Physics and Astrophysics of Quark Gluon Plasma\\
		1-5 February , 2015\\
		Kolkata, India}

\begin{document}

\section{Introduction}
The heavy ion collisions at ultra relativistic energy create matter 
with high energy density required to form Quark Gluon Plasma (QGP).
 Relativistic Heavy Ion Collider (RHIC) and Large Hadron Collider 
(LHC) are designed to create and explore QGP. Many measurements at 
RHIC and LHC already point to the formation of QGP \cite{quarkmatter2014}. 
The heavy quarks (charm and bottom) are produced in hard partonic 
interactions in heavy ion collisions and their initial momentum 
distribution can be calculated from pQCD \cite{kumar2010}.  While 
traversing the hot/dense medium formed in the collisions, these quarks 
loose energy either due to the elastic collisions with the plasma 
constituents or by radiating a gluon or both. There are several 
formulations to calculate collisional 
\cite{braaten1991, peshier2006,   peigne2008, bjorken1982,thoma1991} 
as well as radiative energy loss 
\cite{armesto2004, Armesto:2004vz, glv, dg}. For a review of 
many of these formalism see Ref.~\cite{Jamil2010}. At high parton 
energies, the radiative energy loss becomes much larger than the 
collisional energy loss but at lower energies, these two processes 
can contribute equally with the collisional energy loss being the 
dominant for small values of the parton energy \cite{ducati2007}.

     In this work, we calculate the collisional energy loss of heavy 
quark using Bjorken formalism \cite{bjorken1982, thoma1991} and and 
Peigne and Peshier (PP) formalism \cite{peigne2008}. We calculate the 
radiative energy loss of heavy quarks using reaction operator formalism DGLV 
(Djordjevic, Gyulassy, Levai and Vitev) \cite{glv,dg,wicks} and using 
generalized dead cone approach \cite{Abir:2012pu, Saraswat:2015ena}. We 
calculate the nuclear modification factor including shadowing and energy 
loss for $B$ and $D$ mesons and compare with ALICE and CMS data.

\section{Heavy Quark Production}
The production cross sections of $c \bar c$ and $b \bar b$ pairs are 
calculated to NLO in pQCD using the CT10 parton densities \cite{Lai:2010vv}. 
We use the same set of parameters as that of Ref. \cite{Nelson:2012bc} 
which are obtained by fitting the energy dependence of open heavy flavor 
production to the measured total cross sections. The mass of charm quark 
is taken as 1.29 GeV and the mass of bottom quark is taken as 4.7 GeV. The 
central EPS09 NLO parameter set \cite{Eskola:2009uj} is used to calculate the 
modifications of the parton distribution functions (nPDF) in heavy ion 
collisions, referred as shadowing effects. 

For the fragmentation of heavy quarks into mesons, Peterson fragmentation 
function is used \cite{Peterson:1982ak}.
\begin{equation}
D_{Q}(z) = \frac{N}{z~\Big[1 - \frac{1}{z} - \frac{\epsilon_{Q}}{(1-z)}\Big]^{2}}~.
\end{equation}
Here $z = p^{H}_{T}/p^{Q}_{T}$ and $N$ is the normalization constant which is fixed by 
summing over all hadrons containing heavy quarks~,
\begin{equation}
\sum \int dz~D_{Q}(z) = 1.
\end{equation}
For charm quark :  $\epsilon_{c}$=0.016 and $N$ = 0.2478~~. \\
For bottom quark :  $\epsilon_{b}$=0.0012 and $N$ = 0.05181~~. \\

The distribution peaks at 
\begin{equation}
z_{\rm max}=\frac{(\epsilon_{Q}+2)-\sqrt{\epsilon_{Q}~(\epsilon_{Q}+4)}}{2}.
\end{equation}

\section{Collisional Energy Loss}
In Peigne and Peshier Formalism, the QCD calculation of the rate of energy 
loss of heavy quark per unit distance 
($dE/dx$) in QGP is given by Braaten and Thoma \cite{braaten1991}. 
 Their formalism is an extension of QED calculation of $dE/dx$ for a muon 
\cite{peshier2006} which assumes that the momentum exchange $q \ll E$. 
  Such an assumption is not valid in the domain when the energy of the heavy quark 
$E \gg M^{2}/T$, where $M$ is the mass of the heavy quark. 
  Peigne and Peshier \cite{peigne2008} extended this calculation 
which is valid in the domain $E \gg M^{2}/T$ to give the expression 
for $dE/dx$ as
\begin{equation}
\frac{dE}{dx} = \frac{4 \pi \alpha^{2}_{s} T^{2}}{3}~\Bigg[\Big(1+\frac{N_{f}}{6}\Big)
\log\Big(\frac{E T}{\mu^{2}_{g}}\Big) + \frac{2}{9}\log\frac{E T}{M^{2}} + c(N_{f})\Bigg]~,
\end{equation}
where $T$ is the temperature of QGP medium, $E$ is the energy of heavy quark, 
$N_{f}(=3)$ is the active flavours, $\alpha_{s}(=0.3)$ is the fine structure 
splitting constant and $c(N_{f}) = 0.146 N_{f} + 0.05$. The thermal gluon mass can 
be written as $m_{g}=\mu_{g}/\sqrt{2}$ where 
$\mu_{g} = \sqrt{4~\pi~\alpha_{s}~T^{2}~\Big(1+\frac{N_{f}}{6}\Big)}$ is the Debye 
screening mass. \\
The expression of Bjorken formalism to calculate the collisional energy loss of 
heavy quark is given in Ref. \cite{bjorken1982, thoma1991}

\section{Radiative Energy Loss}
The rate of radiative energy loss of a heavy quark with energy $E$ due to the 
inelastic scattering with the medium is calculated as
\begin{equation}
\frac{dE}{dx} = \frac{<\omega>}{\lambda}~,
\label{presentdEdx}
\end{equation}
where $<\omega>$ is the mean energy of the emitted gluons and $\lambda$ is the 
mean free path length. \\
$<\omega>$ is calculated from generalised dead cone $\mathcal D$ as 
\begin{equation}
<\omega> = \frac{d\omega~\int \mathcal D~ d\eta}{\frac{1}{\omega}~d\omega~
\int \mathcal D~ d\eta},~\mathcal D = \Bigg(1 + \frac{M^{2}}{s}e^{2\eta}\Bigg)^{-2}   
\rm and~ \eta = - \ln\tan\Big(\frac{\theta}{2}\Big)~.
\label{omegaaverage}
\end{equation}
Here $s(=2 E^{2} + 2~E~\sqrt{E^{2}-M^{2}} - M^{2})$ is mandalstam variable and 
$\theta$ is the emission angle.\\
$\lambda$ is calculated as \cite{Abir:2012pu, Saraswat:2015ena}
\begin{equation}
\frac{1}{\lambda} = \rho_{QGP}~\sigma_{2 \rightarrow 3}~.
\label{pathlambda}
\end{equation}
Here $\sigma_{2 \rightarrow 3}$ for the process $2 \rightarrow 3$ is calculated 
as \cite{Biro:1993qt}
\begin{equation}
\sigma_{2 \rightarrow 3} = 4~C_{A}~\alpha^{3}_{s}~\int \frac{1}{q^{2}_{\perp}}dq^{2}_{\perp}~
\int \frac{1}{\omega}~d\omega ~\int~\mathcal D~d \eta.
\label{crosssection23}
\end{equation}
Here $C_{A}=3$ and $q_{\perp}$ is the transverse momentum of the exchanged gluon.\\
Using Eqs. (\ref{presentdEdx}), (\ref{omegaaverage}), (\ref{pathlambda}) and 
(\ref{crosssection23}) and assigning the limits of $q^{2}$, $\omega$ and $\eta$ we get
\begin{equation}
\frac{dE}{dx} = 24~\alpha^{3}_{s}~\rho_{QGP}~\int^{q^{2}_{\perp}|\rm max}_{q^{2}_{\perp}|\rm min}~
\frac{1}{q^{2}_{\perp}}~dq^{2}_{\perp}~\int^{\omega_{\rm max}}_{\omega_{\rm min}}~d\omega ~ 
\int^{\eta_{\rm max}}_{\eta_{\rm min}}~\mathcal D ~d\eta.
\label{dEdxderved}
\end{equation}
Limits are given in Ref.\cite{Abir:2012pu,Saraswat:2015ena}.\\
Equation (\ref{dEdxderved}) is solved to get the following result which we call 
present result
\begin{equation}
\frac{dE}{dx}=24~\alpha^{3}_{s}~\rho_{QGP}~\frac{1}{\mu_{g}}~\Big(1-\beta_{1}\Big)
~\Bigg(\sqrt{\frac{1}{(1-\beta_{1})}~\log\Big(\frac{1}{\beta_{1}}\Big)}-1 \Bigg)
~\mathcal F(\delta)~~.
\end{equation}
Here
\begin{eqnarray}
\mathcal F(\delta) &=& 2\delta-\frac{1}{2}~\log\Bigg(
\frac{1+\frac{M^2}{s}~e^{2\delta}}{1+\frac{M^2}{s}~e^{-2\delta}}\Bigg)-
\Bigg(\frac{\frac{M^2}{s}~\sinh(2\delta)}
{1+2~\frac{M^2}{s}\cosh(2\delta)+\frac{M^4}{s^{2}}}\Bigg)~~,\\ 
\delta &=& \frac{1}{2}~\log\Bigg[\frac{1}{(1-\beta_{1})}~\log\Big(\frac{1}
{\beta_{1}}\Big)~\Bigg(1+\sqrt{1-\frac{(1-\beta_{1})}{\log(\frac{1}{\beta_{1}})}} 
\Bigg)^{2} \Bigg]~, \\
\beta_{1} &=& \mu^{2}_{g}/(C~E~T),\\
C &=& \frac{3}{2} - \frac{M^{2}}{4~ E~T} + 
\frac{M^{4}}{48~E^{2}~T^{2}~\beta_{0}}~
\log\Bigg[\frac{M^{2} + 6~E~T~(1 + \beta_{0})}
{M^{2} + 6~E~T~(1 - \beta_{0})}\Bigg],\\
\beta_{0} &=& \sqrt{1 - \frac{M^{2}}{E^{2}}}~.
\end{eqnarray}
$\rho_{QGP}$ is density of QGP medium. \\
The expression of DGLV formalism to calculate the radiative energy loss of 
heavy quark is given in Ref.\cite{wicks}.

\section{Evolution Model}
The evolution of the system for each centrality bin is governed by an 
isentropic cylindrical expansion with prescription given in 
Ref.~\cite{ZhaoRapp2011}.
 The entropy conservation condition $s(T)\,V(\tau)= s(T_0)\,V(\tau_0)$ 
and equation of state obtained by Lattice QCD along with hadronic resonance 
are used to obtain temperature as a function of proper 
time \cite{vineet2014}. 
 The transverse size $R$ for a given centrality with number of participant $N_{part}$ 
is obtained as  $R(N_{\rm part}) = R_{A} ~ \sqrt{2~A/N_{\rm part}}$,~ where $R_{A}$ is 
radius of the nucleus. The initial entropy density $s(\tau_0)$ is 
\begin{eqnarray}
s(\tau_0)  = {a_{\rm m} \over V(\tau_0)}   \left(\frac{dN}{d\eta} \right) . 
\end{eqnarray}  
 Here $a_m=5$ is a constant which relates the total entropy with the 
multiplicity \cite{Shuryak:1992wc}. The initial volume 
$V(\tau_0) = \pi \left[R(N_{\rm part})\right]^2 \tau_0$ and measured values of 
$dN/d\eta$~for LHC ~\cite{Aamodt:2010cz}  are used for a given centrality. 
The calculated average path length $(L)$ for 0-20 $\%$ centrality is 
5.62 fm and for 0-100 $\%$ centrality is 4.3 fm.

\section{Results and Discussions}
\begin{figure}
\begin{minipage}[t]{7.2cm}
\includegraphics[width=1.10\textwidth]{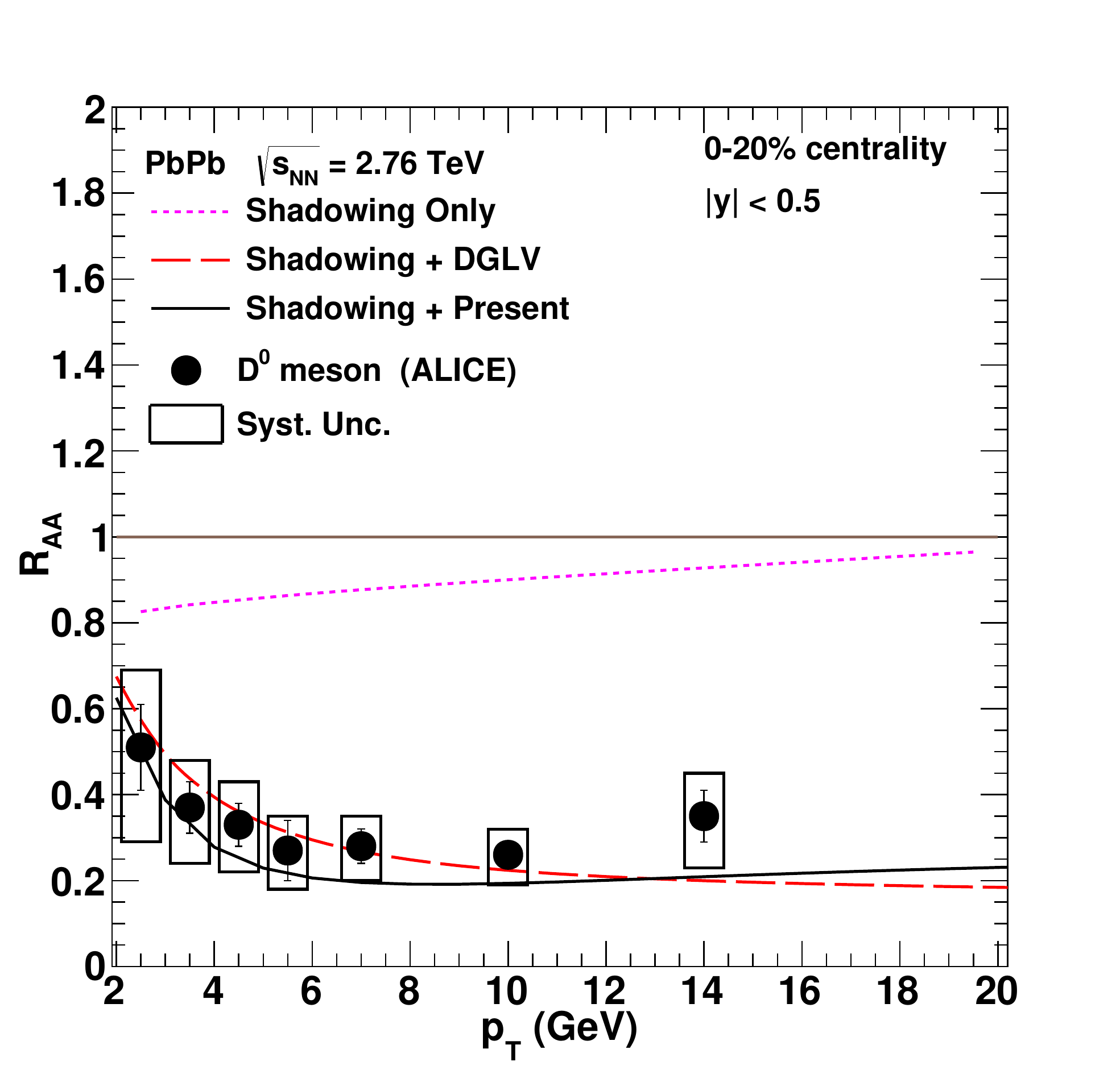}
\caption{Nuclear modification factor $R_{AA}$ of $D^{0}$ mesons 
as a function of transverse momentum using Radiative energy 
loss (DGLV and Present) and shadowing in PbPb collision at 
$\sqrt{s_{NN}}$=2.76 TeV. The data is from ALICE measurements 
of $D^{0}$ mesons~\cite{ALICE:2012ab}.}
\label{alicerad}
\end{minipage}
\hfill
\begin{minipage}[t]{7.2cm}
\includegraphics[width=1.10\textwidth]{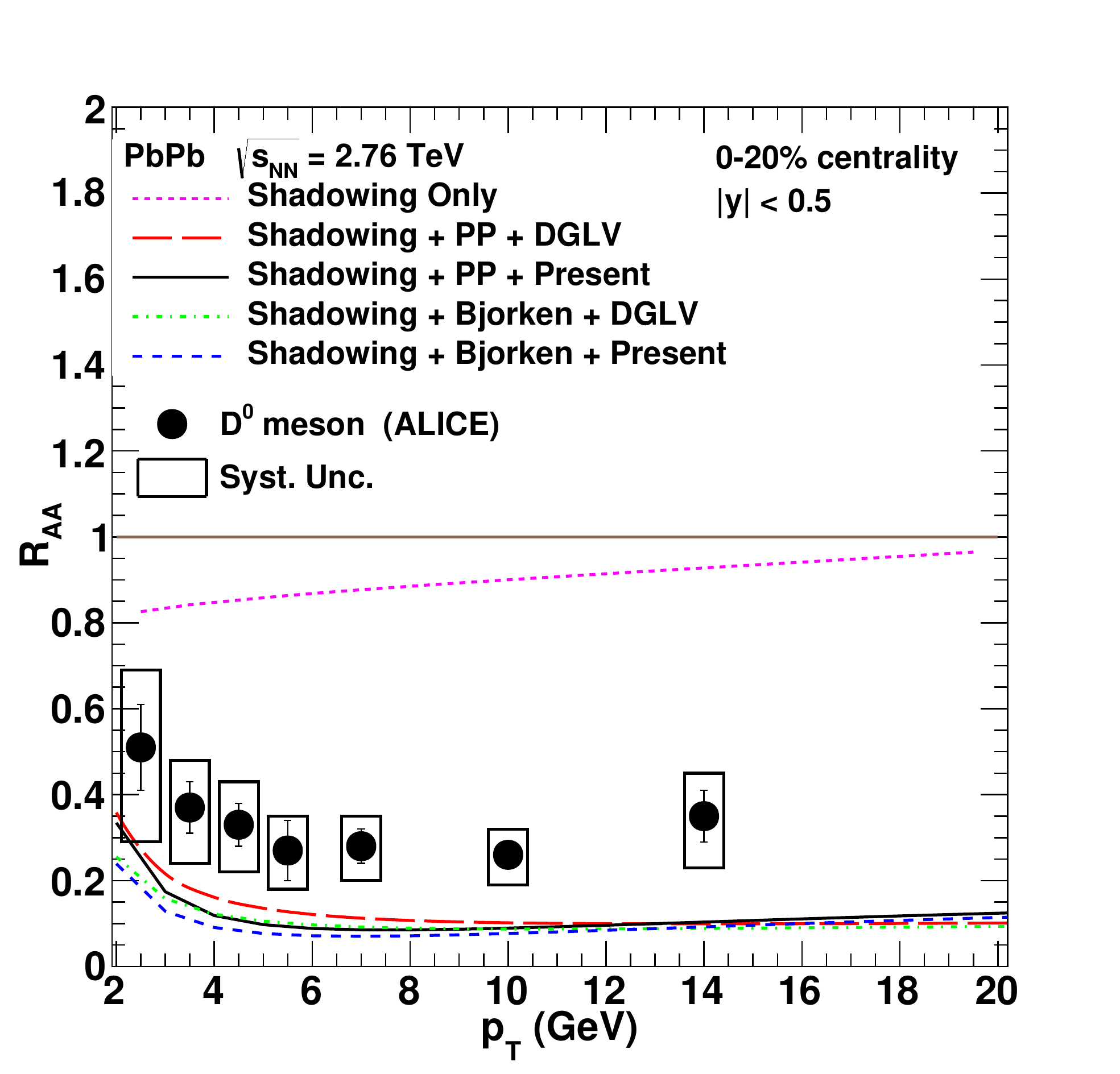}
\caption{Nuclear modification factor $R_{AA}$ of $D^{0}$ mesons 
as a function of transverse momentum using energy loss 
(PP+DGLV and PP+Present) and shadowing in PbPb collision at 
$\sqrt{s_{NN}}$=2.76 TeV. The data is from ALICE measurements 
of $D^{0}$ mesons~\cite{ALICE:2012ab}.}
\label{alicecollrad}
\end{minipage}
\hfill
\end{figure}

Figure \ref{alicerad} shows the nuclear modification factor $R_{AA}$ 
of $D^{0}$ mesons as a function of transverse momentum using Radiative 
energy loss (DGLV and Present) and shadowing in PbPb collision at 
$\sqrt{s_{NN}}$=2.76 TeV. The data is from ALICE measurements 
of $D^{0}$ mesons~\cite{ALICE:2012ab}. The radiative energy loss 
by both DGLV and present calculations explain the data.

Figure \ref{alicecollrad} shows the nuclear modification factor $R_{AA}$ 
of $D^{0}$ mesons as a function of transverse momentum using energy loss 
(PP+DGLV and PP+Present) and shadowing in PbPb collision at 
$\sqrt{s_{NN}}$=2.76 TeV. The data is from ALICE measurements 
of $D^{0}$ mesons~\cite{ALICE:2012ab}. We observe that DGLV+PP, 
DGLV+Bjorken, Present+PP and Present+Bjorken calculations overestimate 
the measured suppression of  $D$ meson.

\begin{figure}
\begin{minipage}[t]{7.2cm}
\includegraphics[width=1.10\textwidth]{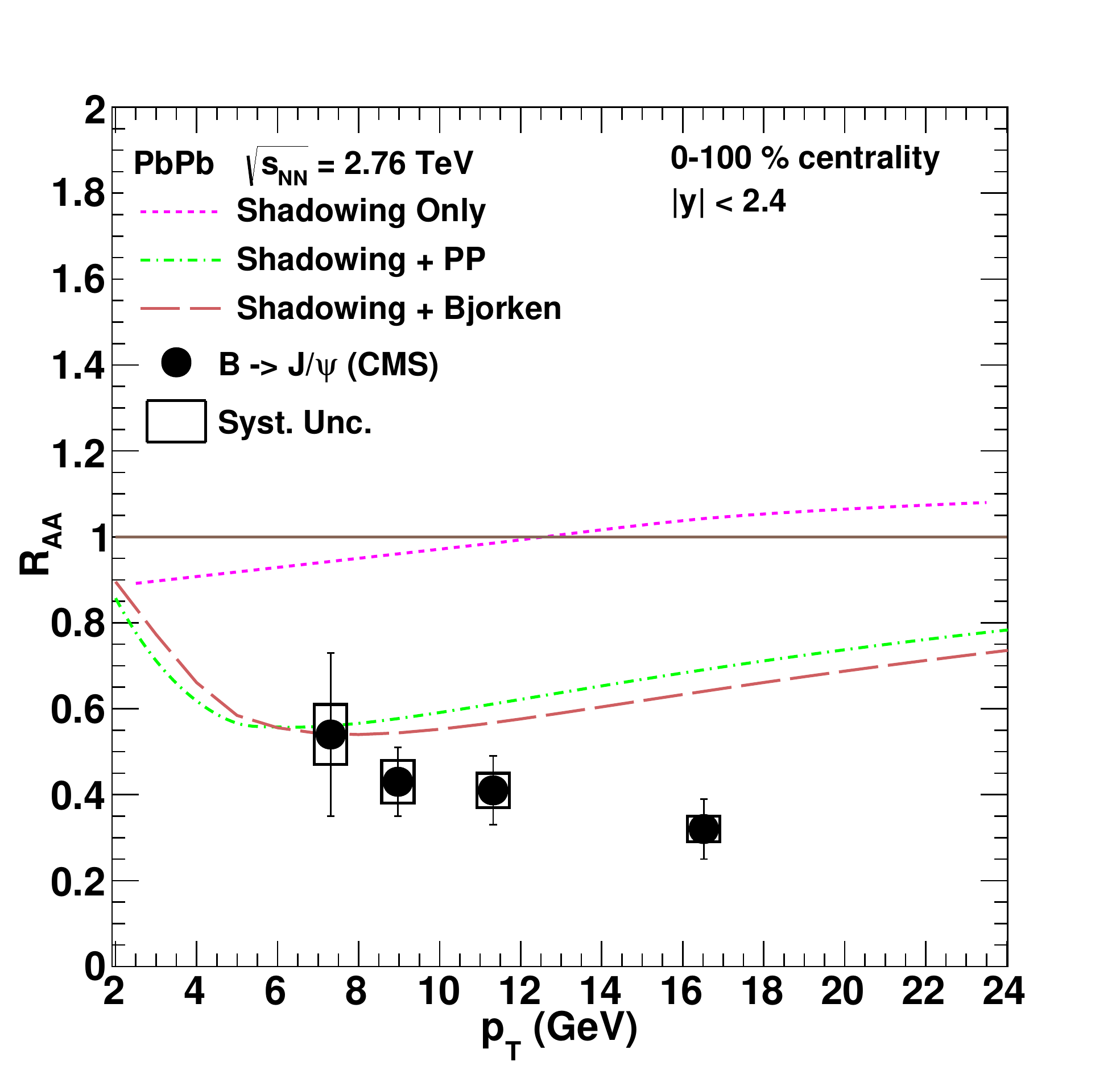}
\caption{Nuclear modification factor $R_{AA}$ of inclusive $J/\psi$
 coming from $B$ mesons as a function of transverse momentum using 
collisional energy loss (PP and Bjorken) and shadowing in PbPb 
collision at $\sqrt{s_{NN}}$=2.76 TeV. The data is from CMS measurements 
of $J/\psi$ mesons from $B$ decay~\cite{cms2014}.}
\label{cmscoll}
\end{minipage}
\hfill
\begin{minipage}[t]{7.2cm}
\includegraphics[width=1.10\textwidth]{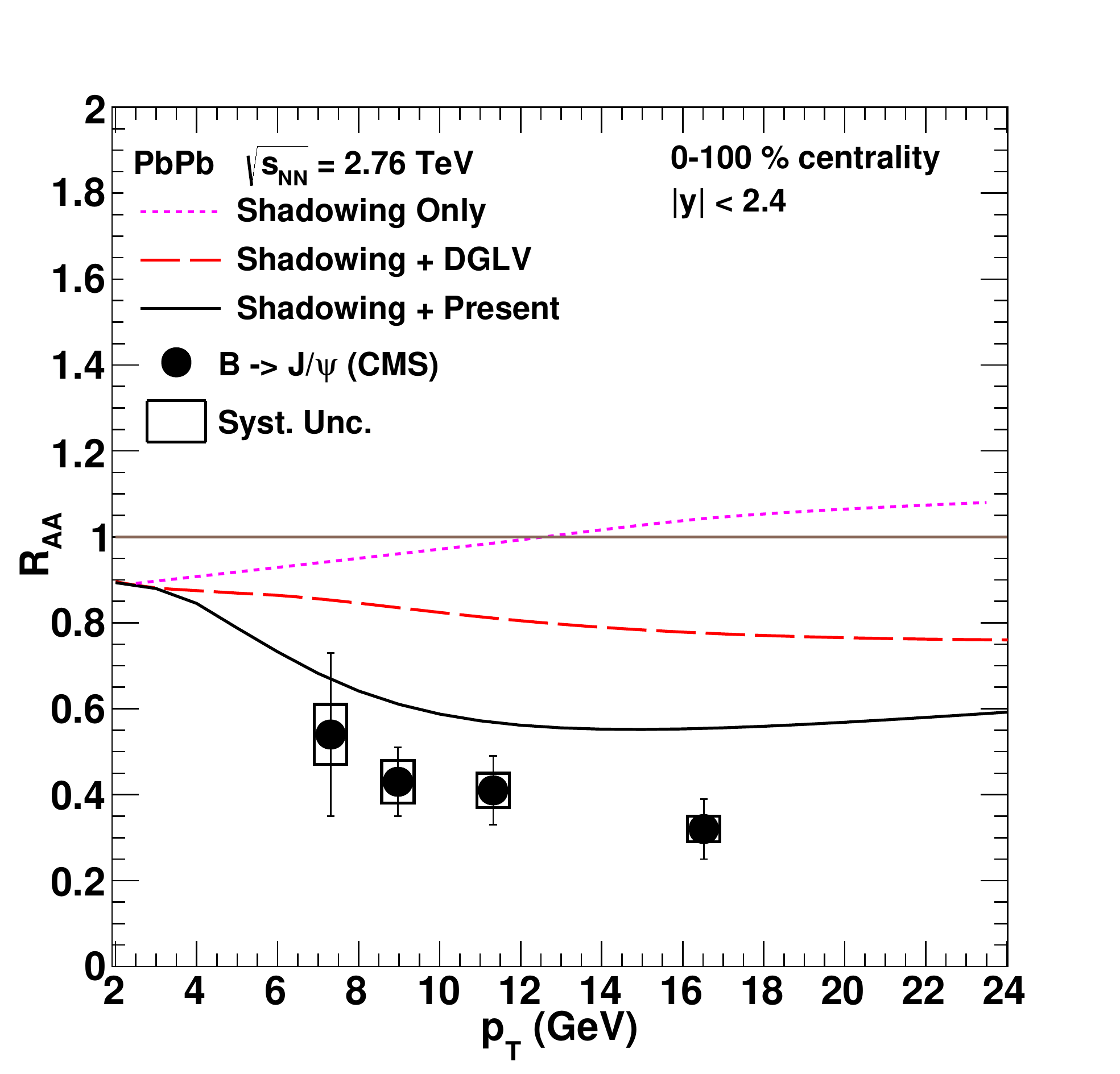}
\caption{Nuclear modification factor $R_{AA}$ of inclusive $J/\psi$ 
coming from $B$ mesons as a function of transverse momentum using 
Radiative energy loss (DGLV and Present) and shadowing in PbPb 
collision at $\sqrt{s_{NN}}$=2.76 TeV. The data is from CMS measurements 
of $J/\psi$ mesons from $B$ decay~\cite{cms2014}.}
\label{cmsrad}
\end{minipage}
\hfill
\end{figure}

Figure \ref{cmscoll} shows the nuclear modification factor $R_{AA}$ of 
inclusive $J/\psi$ coming from $B$ mesons as a function of transverse 
momentum using collisional energy loss (PP and Bjorken) and 
shadowing in PbPb collision at $\sqrt{s_{NN}}$=2.76 TeV. The data is 
from CMS measurements of $J/\psi$ mesons from $B$ decay~\cite{cms2014}.
We observe that the collisional energy loss (PP and Bjorken) calculations 
underestimate the supprression of $B$ meson.

Figure \ref{cmsrad} shows the nuclear modification factor $R_{AA}$ of 
inclusive $J/\psi$ coming from $B$ mesons as a function of transverse 
momentum using radiative energy loss (DGLV and Present) and shadowing 
in PbPb collision at $\sqrt{s_{NN}}$=2.76 TeV. The data is from CMS 
measurements of $J/\psi$ mesons from $B$ decay~\cite{cms2014}. We 
observe that radiative energy loss (DGLV and Present) calculations 
underestimate the suppression of $B$ meson.

\begin{figure}[htp]
\centering
\includegraphics[width=0.60\linewidth]{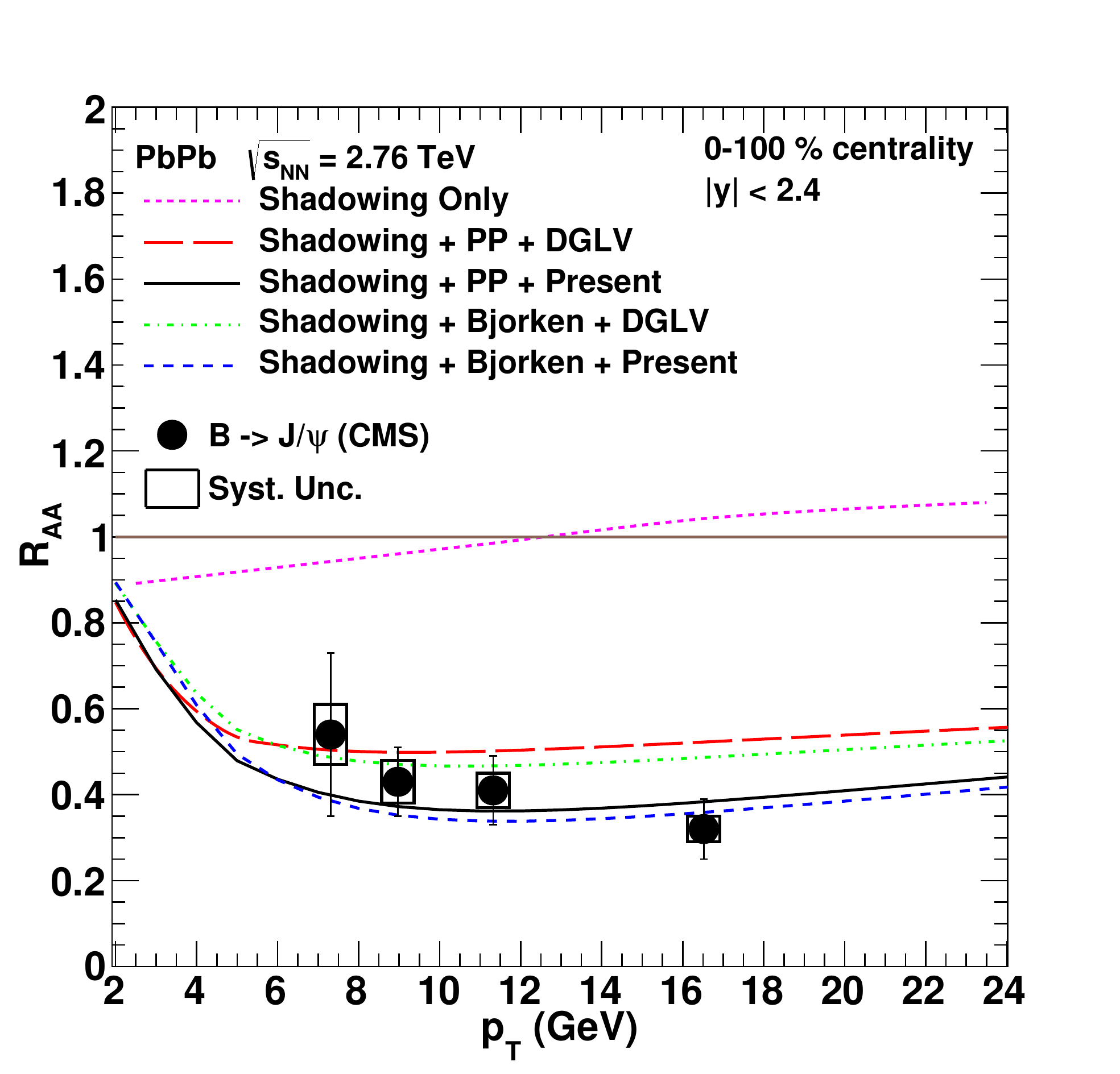}
\caption{Nuclear modification factor $R_{AA}$ of inclusive $J/\psi$ 
coming from $B$ mesons as a function of transverse momentum using 
energy loss (PP+DGLV and PP+Present) and shadowing in PbPb collision at 
$\sqrt{s_{NN}}$=2.76 TeV. The data is from CMS measurements 
of $J/\psi$ mesons from $B$ decay~\cite{cms2014}.}
\label{cmscollrad}
\end{figure}

Figure \ref{cmscollrad} shows the nuclear modification factor $R_{AA}$ 
of inclusive $J/\psi$ coming from $B$ mesons as a function of 
transverse momentum using energy loss (PP+DGLV and PP+Present) and 
shadowing in PbPb collision at $\sqrt{s_{NN}}$=2.76 TeV. The data is 
from CMS measurements of $J/\psi$ mesons from $B$ decay~\cite{cms2014}.
We observe that DGLV+PP and DGLV+Bjorken calculations are consistent with the 
measured suppression of $B$ meson. Present+PP and Present+Bjorken produce 
more suppression than the DGLV+PP but are consistent with the data. 

\section{conclusion}
We study the energy loss of heavy quark due to elastic collisions and gluon 
radiation in hot/dense medium. Results of radiative energy loss are obtained 
from two different formalisms have been compared. Similarly 
the results of collisional energy loss obtained from two different formalisms 
have been compared. The $D$ meson suppression in LHC PbPb collisions can be 
described in terms of radiative energy loss alone but, both the collisional and 
radiative energy loss are required to explain the $B$ meson suppression.

\end{document}